\documentclass[reprint,amsmath,amssymb,aps,pra,longbibliography]{revtex4-1}

\usepackage{float}
\usepackage{graphicx}
\usepackage{epsfig}
\usepackage{epstopdf}
\usepackage{subfigure,color}
\usepackage{dcolumn}
\usepackage{bm}
\usepackage{lineno}
\usepackage{amsmath, amssymb}
\usepackage{accents}
\usepackage{lipsum}
\usepackage{multirow}
\usepackage[usestackEOL]{stackengine}
\usepackage{tabularx}
\usepackage{makecell}
\usepackage{xcolor}
\usepackage{dcolumn}
\usepackage{bm}
\usepackage{booktabs}
\makeatletter
\def\@bibdataout@aps{%
	\immediate\write\@bibdataout{%
		@CONTROL{%
			apsrev41Control%
			\longbibliography@sw{%
				,author="08",editor="1",pages="1",title="0",year="1"%
			}{%
				,author="08",editor="1",pages="1",title="",year="1"%
			}%
		}%
	}%
	\if@filesw \immediate \write \@auxout {\string \citation {apsrev41Control}}\fi 
}
\makeatother

\begin{document}
	
	\preprint{APS/123-QED}
	
	
	

	\title{Independent Control of Multiple Channels in Metasurface Devices}
	
	\author{Xuchen Wang, Ana~D\'{i}az-Rubio, and Sergei A. Tretyakov}
	
	\affiliation{Department of Electronics and Nanoengineering, Aalto University, P.O.~Box 15500, FI-00076 Aalto, Finland}

	\date{\today}

	\begin{abstract}
In analogy with electromagnetic networks which connect multiple input-output ports, metasurfaces can be considered as multi-port devices capable of providing different functionalities for waves of different polarizations illuminating the surface from different directions.
		The main challenge in the design of such multichannel metasurfaces is to ensure independent and full control of the electromagnetic response for each channel ensuring the fulfilment of the boundary condition at the metasurface.
		In this work, we demonstrate that by properly engineering the evanescent fields excited at each port (that is, for all possible illumination directions), it is possible to independently control the reflection or transmission for all 
		different illuminations.
Using the mode-matching method, we analyze the scattering properties of generic space-modulated impedance metasurfaces.
	This method, combined with mathematical optimization, allows us to find a surface impedance profile that simultaneously ensures the desired electromagnetic responses at each port.
		We validate the technique via the design of  phase-controlled multichannel retroreflectors. 
		In addition, we demonstrate that the method is rather powerful in the design of other functional metasurfaces such as multifunctional reflectors and multichannel perfect absorbers.

	\end{abstract}
	
	\maketitle
	
	\section{Introduction} \label{sec:introduction}
Metasurfaces are ultrathin artificial material layers formed by  subwavelength-sized meta-atoms, designed for specific manipulations of the amplitudes, phases, and polarization states of reflected and transmitted  waves \cite{holloway2012overview,kildishev2013planar,yu2014flat,glybovski2016metasurfaces,doi:10.1142/10642-vol1,li2018metasurfaces,bian,yang_rahmat-samii_2019}.
Conventional metasurfaces for control of plane-wave reflection and transmission consist of uniform, periodical arrangements of subwavelength meta-atoms. The period of these  metasurfaces is below the diffraction limit, so that no higher-order propagating modes in free space exist. These metasurfaces obey the usual reflection law.

In the last decade, the concept of metasurfaces has been extended to 
periodical structures whose period exceed half of the wavelength. In this case, multiple diffraction orders are allowed to exist in free space,  and  the metasurface can scatter energy into many directions. 
The traditional design method of such devices is to control the local reflection/transmission phases according to the required phase distributions of scattered fields. This method is based on the phased-array principle, also called the generalized laws of reflection and refraction \cite{yu2011light}.
This approach has been leveraged for the synthesis of ultra-thin optical devices for anomalous reflection and refraction \cite{yu2011light,ni2012broadband,sun2012high}, lensing \cite{aieta2012aberration,chen2012dual}, holographic imaging \cite{ni2013metasurface,yifat2014highly}, and so forth. But those devices  commonly suffer from reduced efficiencies due to impedance mismatch between the incident and diffracted waves \cite{wong2015reflectionless,estakhri2016wave,asadchy2016perfect}.

Recent efforts on metasurface-based gratings have been focused on perfect energy transmission between two scattering channels. Representative works include perfect anomalous refraction  realized with bianisotropic elements \cite{asadchy2016perfect,epstein2018perfect,lavigne2018susceptibility,chen2018theory,kazemi2019simultaneous} and perfect anomalous reflection using the concepts of near-fields engineering \cite{epstein2016synthesis, diaz2017generalized, kwon2018lossless,wong2018perfect}, meta-grating  \cite{ra2017metagratings}, and other alternative means \cite{wong2018perfect,asadchy2016perfect}.
Furthermore, it has been shown that the power carried by a wave incident from one direction can be redistributed between two channels \cite{asadchy2017flat,epstein2016synthesis, epstein2017unveiling} or among arbitrary numbers of diffraction modes \cite{popov2018controlling,popov2019beamforming,rabinovich2019arbitrary}. 

However, in all of those works, metasurfaces are designed only  for one incident angle, and the response for waves  incident from other directions is actually not engineered. Thus, the surface does not provide any functionality for other illuminations. It is worth mentioning that some of these devices can naturally provide multiple functionalities for different incidences \cite{asadchy2017flat,yin2019terahertz}, for example, an anomalous reflector can always serve as a retroreflector when illuminated from its isolated channel \cite{yin2019terahertz}. But those additional functionalities are unconsciously inherited from reciprocity and power conservation.

Recently, interest in multifunctional metasurfaces significantly increased and new devices which perform different functionalities by switching the incident directions \cite{asadchy2017flat,yin2019terahertz} or polarization states \cite{zhang2017multichannel,zhang2019multichannel} have been proposed. 
For example, in \cite{asadchy2017flat}, 
five-ports retroreflectors have been experimentally demonstrated  by engineering the spatial dispersion of a reactive surface.
In \cite{kamali2017angle}, it was demonstrated  that the reflection phases for incidences from two different angles can be independently controlled by exciting different resonant modes in meta-atoms.  
However, these approaches to independent control of channel responses are applicable only for specific functionalities and  strongly rely on numerical optimizations which are time-consuming -- especially when the number of considered channels increases -- and do not provide versatile and systematic design tools for multi-channel devices.

According to the Floquet theory, periodic planar structures scatter into infinite numbers of harmonics in free space, including far-field  propagating modes and near-field evanescent modes.
In the far-zone, propagating modes are equivalent to ``open waveguides'' supporting waves propagating along different directions in space  which are analogous to ``channels'' in a multi-port network \cite{guglielmi1989multimode, asadchy2017flat}. 
Therefore,  metasurfaces can be viewed  as   $N$-port networks, where the numbers of ports depends on the number of considered incidence and scattering directions.
Here, in the spirit of the classical theory of multi-port systems \cite{pozar2009microwave}, we consider linear metasurfaces optimized to operate for plane waves propagating along  $N$ directions in space as multi-port networks  characterized by an $N\times N$ scattering matrix $\overline{\overline{S}}$.
We show that this scattering matrix can be engineered in an arbitrary way provided that the matrix does not violate reciprocity for reciprocal structures  and energy conservation for lossless metasurfaces.
As it is well known from the  electromagnetic networks theory, for lossless and reciprocal metasurfaces the scattering matrix is unitary and symmetric.

The design objective is to implement actual structures that realize a certain $\overline{\overline{S}}$ matrix.
The challenge is to find homogenized surface parameters (such as surface impedances \cite{holloway2012overview}, polarizabilities \cite{bian}, or susceptibilities \cite{achouri2015general}) for multichannel metasurfaces and synthesize meta-atom arrays which  realize the desired functionalities without brute-force numerical optimizations \cite{wang2018extreme}. 
Conventional approaches assume that the  surface properties are  determined by one set of incident, reflected and transmitted waves, and the problem of  finding parameters that can simultaneously satisfy the boundary condition by multiple sets of incident and scattered waves is not addressed. As we show in this paper, the key for solving this problem is to excite proper groups of evanescent modes for incidences from different directions. These engineered evanescent modes together with the desired propagating modes can satisfy one impedance boundary condition for multiple incidence/scattering scenarios. Thus, one metasurface can offer multiple different functionalities for illuminations from different directions.

Our work is based on rigorous analytical formulations for the calculations of scattering harmonics of an arbitrary periodical  space-modulated  metasurface.
Using a simple mathematical optimization, one can synthesise a gradient sheet impedance which can provide the defined multiple functionalities.
We show that the developed semi-analytical method can be used in the design of various types of multichannel metasurfaces, controlling scattering directions, amplitudes, and phases for different illuminations.


\section{Design Concept}

In this work, we use the surface impedance model as the homogenization model for multichannel metasurfaces. 
Surface impedance defines relations between surface-averaged tangential electric and magnetic fields on the metasurface.
After the incident, reflected, and transmitted fields of metasurfaces are ascertained, the required surface impedance can be easily obtained via the current-field relations \cite[\textsection~2.4.3]{yang_rahmat-samii_2019}.
This method has been widely used in the synthesis of metasurfaces modeled by electric impedance sheets \cite{elek2015synthesis,minatti2014modulated,fong2010scalar}, and bianisotrpic metasurfaces realized with cascaded impedance sheets \cite{pfeiffer2014bianisotropic,wong2015reflectionless,epstein2016arbitrary,chen2018theory}.
For a given scattering matrix, the known theoretical methods can engineer only the scattering properties of metasurfaces according to one column of the scattering matrix, and the properties defined in other columns cannot be intentionally engineered. In other words, once the required surface impedance  for excitation of one port (for waves incident from a specific direction) is determined, the responses at other channels are fixed and not controllable. 

One should notice that the $\overline{\overline{S}}$ matrix of a metasurface describes only the reflection and transmission of propagating modes. 
Although evanescent modes are not explicitly present in the scattering matrix of multichannel metasurfaces, they actually offer additional degrees of freedom in the determination of the  boundary condition which can simultaneously provide the desired  properties of the scattering matrix for propagating modes \cite{epstein2016synthesis,asadchy2016perfect}. Based on this observation, we propose the following approach to the design of multichannel metasurfaces: For incidence from one port, we define the scattered propagating modes in the desired $\overline{\overline{S}}$ matrix as well as a set of evanescent harmonics with unknown complex amplitudes;
Likewise, for incidences from other ports, we introduce other sets of evanescent harmonics. In principle, invoking an infinite number of evanescent modes provides complete freedom for finding a surface impedance whose response realizes the desired properties for all channels, that is, creation of multiport metasurface devices with  arbitrarily defined  $\overline{\overline{S}}$ matrices.

The important question is how to determine the required evanescent modes for each incidence scenario.  Analytically solving for the evanescent harmonics is not an easy or even possible task since large, up to infinite numbers of evanescent modes should be considered in sets of nonlinear equations  \cite{kwon2018lossless}.
Solving one specific design problem, the author of  \cite{kwon2018lossless} optimized complex amplitudes of evanescent modes using mathematical tools with the aim to find a locally passive surface impedance for perfect anomalous reflection.
But the developed optimization algorithm is applicable only for lossless and impenetrable boundaries. This limitation does not allow optimization of the surface topology,   since  the effects of evanescent modes inside the metasurface volume are not considered in the impenetrable impedance model.
For this reason, we introduce a more effective optimization method which can find  surface impedance for both lossy and lossless, as well as for both  reflection- and transmission-type multichannel metasurfaces.

Instead of optimizing a large number of evanescent modes, as is done in \cite{kwon2018lossless}, we optimize the Fourier coefficients of the impedance profile until the optimized surface  generates the desired propagating harmonics at multiple incidences. 
The optimized surface launches the proper set of auxiliary evanescent modes, and there is no need to directly optimize them.  
The optimization  method is introduced in Section~\ref{sec: optimization}.

	\section{Theory for calculation of Floquet harmonics}\label{Sec: fundamentals}
	
	In this section, using the mode-matching method 
\cite[\textsection~6.2]{hwang2012periodic}, we develop a general theory for  calculation of scattered harmonics of an arbitrary gradient impedance surface. Based upon this theory, a mathematical optimization method is introduced to control the scattering properties of metasurfaces.
	
	
\subsection{Harmonic analysis for an arbitrary periodic surface impedance} \label{Sec:  analytical model}

\begin{figure}[h!]
		\centering
		\subfigure[]{ 
			\includegraphics[width=0.56\linewidth]{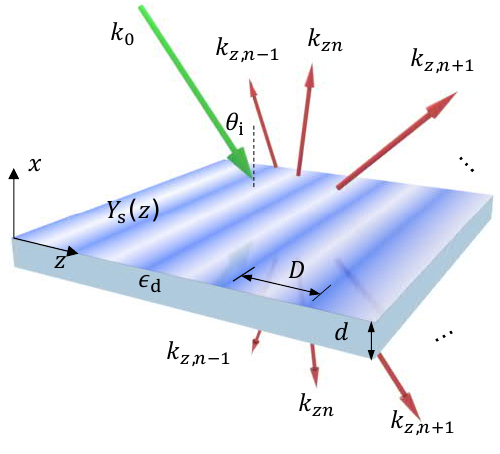}
			\label{fig:scattering_space_modulated}}
		\subfigure[]{ 
			\includegraphics[width=0.37\linewidth]{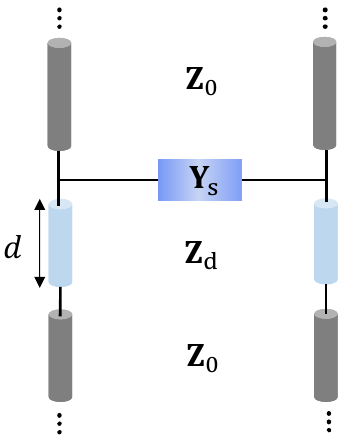}
			\label{fig:space_circuit}}
		\caption{(a) Scattering scenario of a space-modulated surface impedance based on a dielectric substrate. The wave is incident from $\theta=\theta_{\rm i}$; (b) Equivalent circuit of the structure. } \label{fig:analytical_space}
	\end{figure}

Figure~\ref{fig:scattering_space_modulated} illustrates the scattering scenario where a space-modulated impedance sheet $Z_{\rm s} (z)$ is mounted on a dielectric substrate. The sheet impedance is periodically modulated (period $D$) with an arbitrary profile. Under plane-wave excitation, the gradient surface  scatters into an infinite number of Floquet harmonics. The tangential wavevector of the Floquet modes can be written as $k_{zn}=k_0\sin\theta_{\rm i}+n\beta_{\rm M}$, where $\beta_{\rm M}=2\pi/D$ is the spatial frequency of the surface impedance and $n$ is the mode order \cite{asadchy2017flat}. Furthermore, $\theta_{\rm i}$ is the incident angle, and $k_0$ is the free-space wavenumber. Modes satisfying $|k_{zn}|<k_0$ propagate into the far-zone at the angle $\theta_n$, defined by
	\begin{equation}
	\sin\theta_n=-\frac{k_0\sin\theta_{\rm i}+n\beta_{\rm M}}{k_0}.
	\end{equation} 
	If $|k_{zn}|>k_0$, the mode is evanescent with a decaying amplitude along the surface normal.
For a given surface impedance and incident wave, the amplitudes and phases of all the scattered harmonics can be uniquely determined using the mode-matching method \cite[\textsection~6.2]{hwang2012periodic}, by enforcing the boundary condition on the surface. However, this process could be a cumbersome if the structure consists of multiple layers (e.g., with a substrate or superstrate).
	
Here, we model the structure with a simple circuit model, shown in Fig.~\ref{fig:space_circuit}. Similarly with the conventional transmission-line model, the gradient surface impedance is represented by a shunt admittance and the substrate is modeled as a section of a shorted transmission line. 	The tangential electric and magnetic fields on the interfaces are analogized to the voltage and current, respectively.
	Considering that the metasurface scatters infinitely many Floquet modes, the equivalent voltage and current should be also composed of an infinite number of harmonics: 
	\begin{subequations}
		\begin{equation}
		I_{\rm s}(z)=\sum_{n=+\infty}^{-\infty}i_{\rm s}^n e^{-jk_{zn}z},  \label{Eq: current_space_modulation}
		\end{equation} 
		and
		\begin{equation}
		\quad V_{\rm s}(z)=\sum_{n=+\infty}^{-\infty}v_{\rm s}^n e^{-jk_{zn}z}.\label{Eq: voltage_space_modulation}
		\end{equation}
	\end{subequations}
The complex amplitudes of voltage and current harmonics can be written in the form of vectors, $\accentset{\rightharpoonup}{v}_{\rm s}=[\cdots,v_{\rm s}^{-1},v_{\rm s}^{0}, v_{\rm s}^{+1}, \cdots]^T$ and $\accentset{\rightharpoonup}{i}_{\rm s}=[\cdots,i_{\rm s}^{-1},i_{\rm s}^{0}, i_{\rm s}^{+1}, \cdots]^T$. The quantity which relates the current and voltage vectors is called admittance matrix, $\accentset{\rightharpoonup}{i}_{\rm s}={\bf Y}_{\rm s}\cdot \accentset{\rightharpoonup} {v}_{\rm s}$. Next, we aim to find the admittance matrix of  gradient impedance surfaces.
	
	 Due to periodicity of the surface, the surface admittance $Y_{\rm s}(z)=1/Z_{\rm s}(z)$ can be expanded into Fourier series:
	\begin{equation}
	Y_{\rm s}(z)=\sum_{m=-\infty}^{+\infty}g_{m} e^{-jm\beta_{\rm M} z}.
	\label{Eq: Fourier_series_grid_impedance}
	\end{equation}
	Substituting Eqs.(\ref{Eq: current_space_modulation}), (\ref{Eq: voltage_space_modulation}), and (\ref{Eq: Fourier_series_grid_impedance}) into  Ohm's law $I_{\rm s}(z)=Y_{\rm s}(z)V_{\rm s}(z)$, we obtain equation 
	\begin{equation}
	\sum_{n=-\infty}^{+\infty}i_{\rm s}^n e^{-jk_{zn} z}=\sum_{\ell=-\infty}^{+\infty}\sum_{m=-\infty}^{+\infty}g_mv_{\rm s}^{\ell} e^{-jk_{z,\ell+m} z}. \label{Eq:admittance_expansion}
	\end{equation}
	By replacing $\ell$ with $\ell-m$ in the right side of Eq.~(\ref{Eq:admittance_expansion}), both sides have the same basis functions. In this case, the expression for the current of each harmonic can be simplified as
	\begin{equation}
	i_{\rm s}^n=\sum_{m=-\infty}^{+\infty}g_m v_{\rm s}^{n-m}.\label{Eq: current_voltage_admittance}
	\end{equation}
It can be seen from Eq.~(\ref{Eq: current_voltage_admittance}) that the $n$-th current component is coupled with all the voltage harmonics. Considering a finite number of modes (from  $-N$ to $+N$),  we can list $(2N+1)$ harmonics that will contribute  in  Eq.~(\ref{Eq: current_voltage_admittance}). This relation  can be written in a matrix form:
	\begin{equation}
	\begin{pmatrix}
	i_{\rm s}^{-N}  \\
	i_{\rm s}^{1-N} \\
	\vdots   \\
	i_{\rm s}^{+N} \end{pmatrix}=\begin{pmatrix}
	g_{0} & g_{-1} & \cdots & g_{-2N} \\
	g_{1} & g_{0} & \cdots & g_{1-2N} \\
	\vdots  & \vdots  & \ddots & \vdots  \\
	g_{2N} & g_{2N-1} & \cdots & g_{0}
	\end{pmatrix}\begin{pmatrix}
	v_{\rm s}^{-N}  \\
v_{\rm s}^{1-N} \\
	\vdots   \\
	v_{\rm s}^{+N} \end{pmatrix}.
	\end{equation}
We can see that the current and voltage vectors are associated with a Toeplitz matrix ${\bf Y}_{\rm s}$ which  we call \emph{admittance matrix}. The admittance matrix is determined only  by the Fourier coefficients of the modulation function and  filled with  ${\bf Y}_{\rm s}(r,c)=g_{r-c}$ at the $r$-th row and $c$-th column. Similarly with scalar admittance \cite{pozar2009microwave}, 
	the transfer matrix (also known as $ABCD$ matrix) corresponding to ${\bf Y}_{\rm s}$ reads 
	\begin{equation}
	{\bf T}_{\rm s}=\begin{bmatrix}
	{\bf I}&{\bf 0}\\
	{\bf Y}_{\rm s}&{\bf I}
	\end{bmatrix},
	\end{equation}
	where $\bf I$ is the $(2N+1)\times(2N+1)$ identity matrix.
	
Next we derive the transfer matrix of the dielectric slab ${\bf T}_{\rm d}$. Since no modulation is  assumed in dielectric layers, there is no coupling terms (off-diagonal terms) in the characteristic impedance matrix of dielectric ${\bf Z}_{\rm d}$, and the matrix has only  diagonal terms ${\bf Z}_{\rm d}(n,n)$ (the row and column of the matrix are indexed from $-N$ to $+N$), where  ${\bf Z}_{\rm d}(n,n)$ is the characteristic impedance of the $n$-th harmonic. 
	The characteristic impedances for different polarization states are different and are expressed as
	\begin{equation}
	{\bf Z}_{\rm d}^{\rm TM}(n,n)=\frac{k_{xn}^{\rm d}}{\omega_0\epsilon_0\epsilon_{\rm d}} \quad\mathrm{and} \quad {\bf Z}_{\rm d}^{\rm TE}(n,n)=\frac{\mu_0\omega_0}{k_{xn}^{\rm d}}, \label{Eq: wave_impedance_TETM}
	\end{equation}	
where $k_{xn}^{\rm d}=\sqrt{\omega_0^2\epsilon_0\epsilon_{\rm d}\mu_0-k_{zn}^2}$ is the normal component of the  wavevector in dielectric. 
Due to the finite thickness of substrates, their transfer matrices must include phase delay  \cite{pozar2009microwave}. In this multi-mode system, we define  propagation matrix ${\bf P}_{\rm d}$, which also contains only diagonal terms ${\bf P}_{\rm d}(n,n)=e^{-jk_{xn}^{\rm d} d}$, modeling phase delay of each harmonic in dielectric slabs. 
	The transfer matrix  of dielectric layers can be expressed as \cite{hadad2016breaking} 
	\begin{equation}
	{\bf T}_{\rm d}=
	\begin{bmatrix}
	({\bf P}_{\rm d}+{\bf P}_{\rm d}^{-1})/2&-{\bf Z}_{\rm d}({\bf P}_{\rm d}-{\bf P}_{\rm d}^{-1})/2\\
	-{\bf Z}_{\rm d}^{-1}({\bf P}_{\rm d}-{\bf P}_{\rm d}^{-1})/2&({\bf P}_{\rm d}+{\bf P}_{\rm d}^{-1})/2
	\end{bmatrix}.
	\end{equation}
	
It is necessary to mention that the metasurface structure can contain a cascade of modulated  impedance sheets (with the same periodicity $D$) separated with dielectric substrates. 
Knowing the transfer matrices of all  constitutive layers, one can simply multiply them in sequence from the first illuminated layer, 
	\begin{equation}
	{\bf T}_{\rm tot}= \cdots{\bf T}_{\rm s}{\bf T}_{\rm d}\cdots= \begin{bmatrix}
	{\bf A}&{\bf B} \\
	{\bf C}&{\bf D} 
	\end{bmatrix}, \label{Eq: casade}
	\end{equation}
where dots represent additional constitutive layers for metasurfaces with multiple  impedance sheets  or dielectric layers. 
For impenetrable metasurfaces in which the transmission is blocked by a metallic plate, the transfer matrix  can be obtained by multiplying the transfer matrix by that of a  metal plate in the end of Eq. (\ref{Eq: casade}).
	
The incident, reflected and transmitted harmonics are written as vectors (denoted as $\accentset{\rightharpoonup} {v}_{\rm in}$, $\accentset{\rightharpoonup} {v}_{\rm re}$ and $\accentset{\rightharpoonup} {v}_{\rm tr}$, respectively) and are related by the total transfer matrix
	\begin{equation}
	\begin{bmatrix}
	\accentset{\rightharpoonup} {v}_{\rm in}+\accentset{\rightharpoonup} {v}_{\rm re} \\
	{\bf Y}_0\cdot(\accentset{\rightharpoonup} {v}_{\rm in}-\accentset{\rightharpoonup} {v}_{\rm re})
	\end{bmatrix}
	=
	\begin{bmatrix}
	{\bf A}&{\bf B} \\
	{\bf C}&{\bf D} 
	\end{bmatrix}
	\begin{bmatrix}
	\accentset{\rightharpoonup} {v}_{\rm tr} \\
	{\bf Y}_0 \cdot \accentset{\rightharpoonup} {v}_{\rm tr}
	\end{bmatrix}, \label{Eq: ABCD_space_modulated}
	\end{equation}
	where ${\bf Y}_0={\bf Z}_0^{-1}$ is the admittance matrix of free space (${\bf Z}_0$ has the same format with ${\bf Z}_{\rm d}$ only by replacing $\epsilon_{\rm d}$ in Eq.~(\ref{Eq: wave_impedance_TETM}) with 1). We define reflection and transmission matrices that allow us to calculate  reflected and transmitted harmonics for a given incident wave: $\accentset{\rightharpoonup} {v}_{\rm re}={\bf \Gamma}\cdot\accentset{\rightharpoonup} {v}_{\rm in}$ and $\accentset{\rightharpoonup} {v}_{\rm tr}={\bf T}\cdot\accentset{\rightharpoonup} {v}_{\rm in}$. From Eq. (\ref{Eq: ABCD_space_modulated}), the transmission and reflection matrix are calculated as
	\begin{subequations}
		\begin{equation}
		{\bf T}=2\left(	{\bf A}+
		{\bf B}{\bf Y}_0+{\bf Z}_0{\bf C}+
		{\bf Z}_0{\bf D}{\bf Y}_0
		\right)^{-1},
		\end{equation}
		and
		\begin{equation}
		{\bf \Gamma}=\left({\bf A}+
		{\bf B}{\bf Y}_0\right) 	{\bf T}- 	{\bf I}.\label{Eq: reflection matrix}
		\end{equation}\label{Eq: reflection and transmission}
	\end{subequations}
We note that $\bf T$ and $\bf \Gamma$ are $(2N+1)\times(2N+1)$ square matrices. The columns and rows of $\bf T$ and $\bf \Gamma$ are indexed from $-N$ to $+N$. Likewise, the elements in $\accentset{\rightharpoonup} {v}_{\rm in}$, $\accentset{\rightharpoonup} {v}_{\rm re}$ and $\accentset{\rightharpoonup} {v}_{\rm tr}$ are also indexed from $-N$ to $+N$. 
For illumination from one specific direction, the $0$-th element of the incidence vector is $1$ while the rest positions are filled with zeros,  $\accentset{\rightharpoonup} {v}_{\rm in}=[\cdots,0,1,0,\cdots]^T$. The multiplication results of the incidence vector with the reflection or transmission matrices is actually the $0$-th column of  $\bf T$ or  $\bf \Gamma$. Therefore, the $n$-th order of reflected and transmitted harmonics can be found as ${v}_{\rm re}^n={\bf \Gamma}(n,0)$ and ${v}_{\rm tr}^n={\bf T}(n,0)$, respectively.

\subsection{Engineering scattering matrix by mathematical optimization} \label{sec: optimization}
	
The presented analytical  method is rigorous and general.  For an incident plane wave with an  arbitrary polarization state  and incident angle, the scattered harmonics (including propagating and evanescent modes) are uniquely determined by  the Fourier coefficients $g_m$ which define the transfer matrix and the reflection and transmission coefficients (\ref{Eq: reflection and transmission}).  
	On the other hand, if we expect the occurrence of some specific harmonics, e.g., propagating  modes defined in  $S$ matrix of multichannel metasurfaces,  it is possible to  find  a proper set of $g_m$ that ensure excitation of the desired harmonics. 
	However, the analytical relations between $S_{ij}$ and $g_m$ are not straightforward due to the need of complicated  matrix operations. For this reason, in practice it is not possible to analytically solve $g_m$ for desired harmonics. 
	
	Here, we use  mathematical optimization tools available in MATLAB package to find the optimal values of $g_m$. Each element of the desired scattering matrix is an objective function in optimization. We denote the $k$-th objective as $O_k$ which can be the amplitude or phase of one scattering parameter $S_{ij}$. In each trial of the optimization, MATLAB assumes an array of $g_m$ and calculates the realized value $F_k$ for each objective $O_k$, using the analytical formulas in Section~\ref{Sec:  analytical model}.
	For a system with multiple objectives, we can define a cost function in the optimization code as 
	\begin{equation}
	C=\sum_{k}\frac{|F_k-O_k|}{|O_k|}.\label{Eq: cost function}
	\end{equation}
The cost function $C$ is a sum of normalized errors calculated for all the objectives.
Employing \textit{MultiStart} and \textit{fmincon} optimization algorithms, MATLAB can search for the minimum value of $C$ in the multidimensional parameter space.
The optimization tools also allow us to set constraints on the Fourier coefficients in order to satisfy specific additional conditions, e.g., lossless, passive, capacitive or inductive impedance. 
It is worth to mention that, normally, the number of unknowns $g_m$ should be larger than the numbers of objectives. Introducing more $g_m$ increases the possibilities to find the optimal solution, but on the other hand, it may increase the optimization time. A similar optimization approach has been employed to determine the cascaded tensor admittances in the synthesis of bianisotropic metasurfaces \cite{pfeiffer2014bianisotropic}.
In Section~\ref{Sec: phase controlled},  we present the detailed process of optimization of the surface impedance for multichannel retroreflectors with independently and arbitrarily controlled reflection phases. In addition, in  Section~\ref{Sec: other possibilities} we use the same optimization methods for the design of multifunctional reflectors and multichannel perfect absorbers.

	\section{Phase-controlled  multichannel retroreflectors}\label{Sec: phase controlled}
	
	\subsection{Conventional retroreflectors}\label{Sec: conventional retroreflectors}
\begin{figure}[h!]
		\centering
		\includegraphics[width=0.9\linewidth]{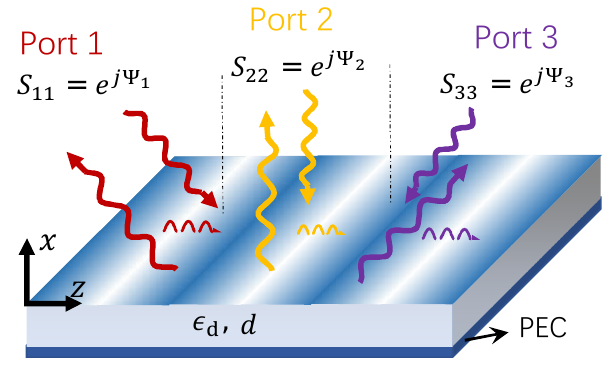}
		\caption{Schematics for phase controlled multichannel metasurfaces. }\label{fig:schemetic}
	\end{figure}

	Let us first recall the design idea of a conventional three-channel retroreflector  presented in \cite{asadchy2017flat}. A lossless impedance surface is positioned on a grounded substrate, as shown in Fig.~\ref{fig:schemetic}. The period of impedance sheet modulation is set to $D=\lambda/2\sin\theta_{\rm i}$, where $\theta_{\rm i}$ is the incident angle.  The operating channels are defined at  $\theta=+\theta_{\rm i},\theta=0$, and $\theta= -\theta_{\rm i}$, corresponding to Ports 1, 2, and 3, respectively. 
	
	The periodicity of the system  allows only specular and retroreflection when the metasurface is illuminated by plane waves at $\theta=\pm\theta_{\rm i}$.
	Since the period is subwavelength  (for $30^\circ<\theta_{\rm i}<90^\circ$), waves incident from Port~2 are always fully reflected back because waves to other channels are not allowed to propagate. In this case, if complete retroreflection is ensured for incidence from Port~1, full retro reflectance from Port~3 is automatically satisfied due to reciprocity.
	To design such metasurfaces, we only need to define the incident and reflected fields (both amplitudes and phases) for Port~1, and find the required surface impedance. 
	Although the reflection phase of Port 1 can be arbitrary defined, at other ports, the reflection phases are not controllable.
	\begin{figure}[h!]
		\centering
		\includegraphics[width=0.75\linewidth]{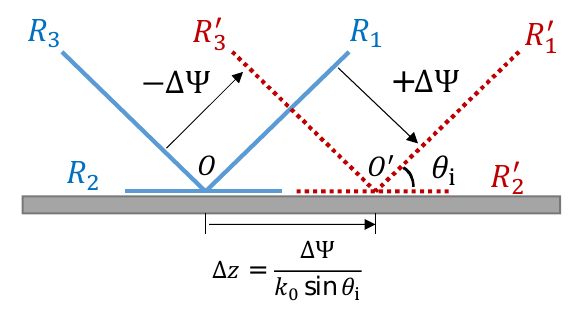}
		\caption{Reference planes when moving the reference point from $O$ to $O^\prime$. $R_1$, $R_2$, and $R_3$ are the reference planes for Ports 1, 2, and 3, respectively, when the reference point is selected at $O$.
			$R_1^\prime$, $R_2^\prime$, and $R_3^\prime$ are the reference planes for Ports 1, 2, and 3 when the reference point is selected at $O^\prime$. }\label{fig:reference_plane}
	\end{figure}

	Here, it is very necessary to clarify the definition of reflection phases of multichannel retroreflectors.	
	The reflection phases depend on  the chosen reference planes. For normal incidence (Port~2), it is clear that the reference plane is defined at the metasurface plane ($R_2$ or $R_2^\prime$ plane in Fig.~\ref{fig:reference_plane}). 
	But for oblique incidence (Port~1 or 3), it is necessary to first specify a reference point on the metasurface plane. The reference point ($O$ or $O^\prime$ in Fig.~\ref{fig:reference_plane}) is the intersection point of the  reference plane and the metasurface plane. Apparently, the reflection phase of a retroreflector depends on  selection  of the reference point. 
	However, for  metasurface retroreflectors, the sum of reflection phases from Ports~1 and 3 is always a fixed value, no matter at which position the reference point is located. 
	As shown in Fig.~\ref{fig:reference_plane}, when moving the reference point from $O$ to $O^\prime$, the reference plane of Port~1 moves from $R_1$ to $R_1^\prime$ and the reflection phase increases by $\Delta\Psi$, but for Port~3, the reference plane moves from $R_3$ to $R_3^\prime$, and the phase decreases by the same value $\Delta\Psi$. Therefore, the sum of them is not dependent on the selection of the reference point. 
	In other words, for a specific design, the reflection phases of Ports~1 and 3 can linearly decrease or increase in the  opposite way when moving the reference point along the surface.

	\subsection{Multichannel control of reflection phase}	
	
	The scattering matrix of a  three-channel lossless retroreflector can be written as 
	\begin{equation}
	\overline{\overline{S}}=
	\begin{bmatrix}
	e^{j\Psi_1}&0&0\\
	0&e^{j\Psi_2}&0\\
	0&0&e^{j\Psi_3}
	\end{bmatrix}.
	\end{equation}
Physically, the reflection phases from three ports can be arbitrary  because the matrix is always symmetric and unitary.
	Therefore, it is possible to arbitrarily and independently engineer the reflection phases for incidence from each port.

The reflection matrix  of metasurface $\mathbf \Gamma$ calculated from Eq. (\ref{Eq: reflection matrix}) depends on the direction of incidence, i.e., we can define three different reflection matrices ($\mathbf \Gamma_1$, $\mathbf \Gamma_2$, and $\mathbf \Gamma_3$) for illuminations from Ports 1, 2, and 3.
Our goal is to realize retroreflection at each port with a specific set of reflection phases $\Psi_1$, $\Psi_2$, and $\Psi_3$.  In this scenario, the cost function defined in Eq. (\ref{Eq: cost function}) contains three objectives. 
First, we should ensure  perfect retroreflection from Port 1 (order of $n=-1$), that is, ensure that the amplitude of the retroreflected wave is unity. This condition can be mathematically expressed as
\begin{equation}
	F_1=\left|{\bf \Gamma}_1 (-1,0)\right|, \quad O_1=|S_{11}|=1.
	\label{Eq: objectives_function_port1}
\end{equation}
As it was explained in Section \ref{Sec: conventional retroreflectors}, perfect retroreflection at Port 1 automatically ensures the same functionality at Port 2 and Port 3. 
The second objective is that the reflection phase from Port 2 (order  $n=0$) should be $\Psi_2$:
	\begin{equation}
	F_2=\angle {\bf \Gamma}_2(0,0), \quad O_2=\angle S_{22}=\Psi_2.
	\label{Eq: objectives_function_port2}
	\end{equation}
Finally, the last objective is that the sum of reflection phases from Port 1 and Port 3  has a fixed value $\Psi_{13}=\Psi_1+\Psi_3$:
	\begin{subequations}
		\begin{equation}
		F_3=\angle {\bf \Gamma_1} (-1,0)+\angle {\bf \Gamma_3} (1,0), \label{Eq: objectives_function_port13}
		\end{equation}
		\begin{equation}
			O_3=\angle S_{11}+\angle S_{33}=\Psi_{13}.
		\end{equation}
	\end{subequations}	
At this point, it is important to mention that optimizing the sum of $\Psi_1$ and $\Psi_3$  is more efficient than individually optimize them, because the number of objectives is reduced. 
As we discussed in Section~\ref{Sec: conventional retroreflectors}, once the sum of the reflection phases is fixed, the desired reflection phases of Port 1 and Port 3 can be realized by moving the reference point along the metasurface plane.

It is easy to see that functions $F_k$  depend on coefficients $g_m$. Consequently, optimization will allow us to  search for solutions of $g_m$ that can simultaneously satisfy equations $|F_k-O_k|=0$ ($k=1,2,3$). In this particular example, there are three  equations,  therefore we should introduce more than three coefficients $g_m$ in the surface admittance expansion to find a solution.  Moreover, in order to ensure that $Y_{\rm s}$ is a purely imaginary number (lossless metasurface), we  need to take into consideration that $g_m$ and $g_{-m}$ are not independent and satisfy relations $\Re(g_m)+\Re(g_{-m})=0$ and $\Im(g_m)=\Im(g_{-m})$.

\begin{figure}[h!]
		\centering
		\includegraphics[width=0.9\linewidth]{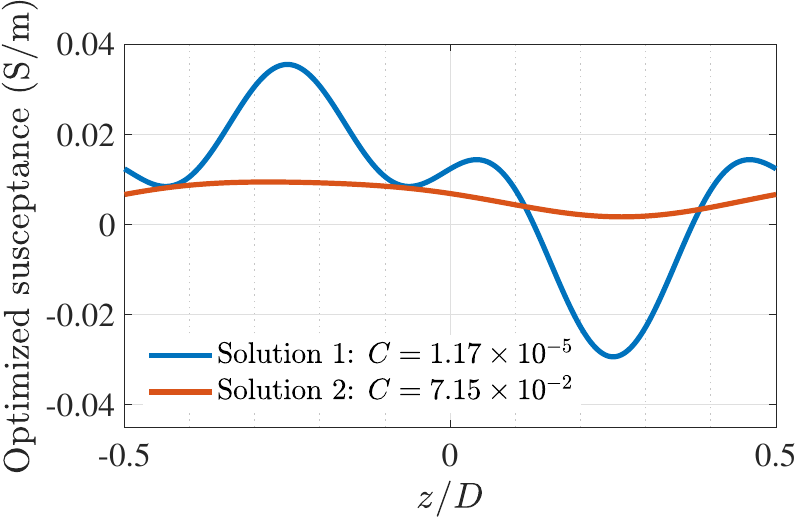}
		\caption{Optimized grid susceptance profiles for desired phase responses at three ports: $\Psi_1=0$, $\Psi_2=-\pi/3$, and $\Psi_3=\pi$. The optimized Fourier amplitudes for Solution~1 read $g_0=7.7\times10^{-3}$, $g_1=-1.05j\times10^{-2}$, $g_2=5.73j\times10^{-3}$ and $g_3=0$. For Solution~2,  $g_0=6.15\times10^{-3}$, $g_1=(5.62-192.26j)\times10^{-5}$, $g_2=(3+1.036j)\times10^{-4}$ and $g_3=0$. The substrate is chosen as $d=0.215$mm  and $\epsilon_{\rm d}=4.2$. The operating frequency is $f=75$~GHz. The incidence is assumed to be TE-polarized. }\label{fig:optimization solution}
	\end{figure}

As an example, we realize a three-channel retroreflector with reflection phases $\Psi_1=0$, $\Psi_2=-\pi/3$, and $\Psi_3=\pi$.  
We introduce four complex unknowns $g_0$, $g_{\pm1}$, $g_{\pm2}$, and $g_{\pm3}$ 
and assume that the other Fourier coefficients are zero. 
	The optimization results show that there may exist multiple solutions which  minimize the cost function.  For example, Fig.~\ref{fig:optimization solution} shows two typical solutions.  In Solution~1, the  surface admittance in one unit cell exhibits  both capacitive and inductive properties at different positions along the $z$-axis, so this function crosses zero. This indicates that the required surface impedance contains some extreme values (zero admittance corresponds to infinite impedance) which may be difficult to implement by patterning a thin electric film.
	In this situation, we can impose additional constraints on $g_m$ to ensure that the surface reactance is always capacitive or inductive along the metasurface, as it is shown in Solution~2. 

	\begin{figure}[h!]
		\centering
		\includegraphics[width=1.0\linewidth]{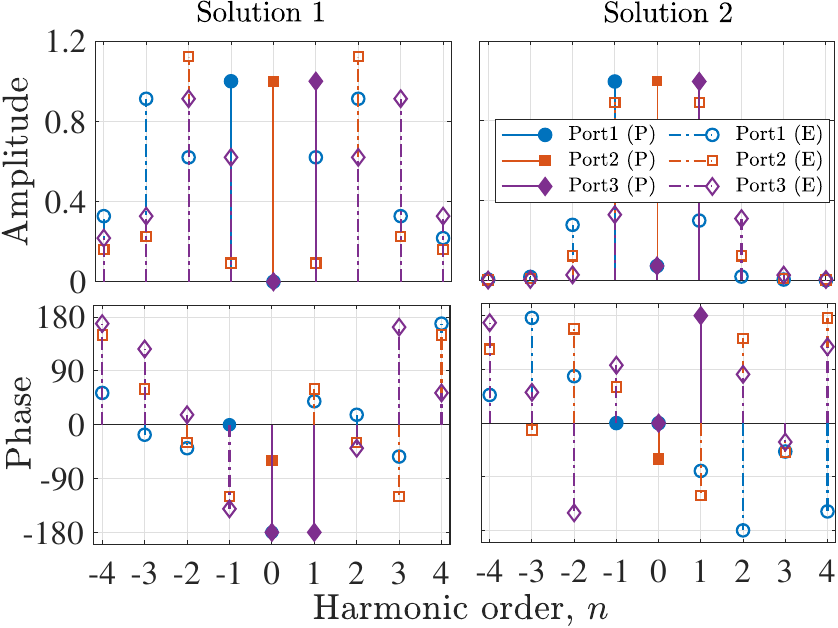}
		\caption{Amplitudes and phases of Fourier harmonics for the optimized solutions. `P' and `E' in the legend represent for propagating and evanescent modes. }\label{fig:harmonics for optimized surface impedance}
	\end{figure}
	Figure~\ref{fig:harmonics for optimized surface impedance} shows the calculated amplitudes and phases of the scattered harmonics for incidences from the three ports, for the two optimized admittance profiles. We can see that the two different solutions  produce almost the same far fields for illuminations from each port, but the excited evanescent modes are totally different. The near fields affect the quality factor of the  device (strong near fields mean a large quality factor and narrow bandwidth).

	\begin{table}[tb]
		\centering
		\caption{\label{tab:1}Optimized amplitudes of the Fourier harmonics for different values of the desired $\Psi_2$. In all the cases, $\Psi_1+\Psi_3=\pi$. }
		\resizebox{0.48\textwidth}{!}{%
			\begin{tabular}{|c| c| c| c| c|c|} 
				\toprule
				\makecell{$\Psi_2$} & \makecell{ $g_0$ } & \makecell{ $g_1$} & \makecell{ $g_2$ }& \makecell{ $g_3$ } & \makecell{  Opt. Err. } \\  \hline
				\hline
				$-\pi$&$j1.88$ $\times10^{-2}$  & $j4.11$ $\times10^{-5}$ & $-j2.28$ $\times10^{-4}$ & $j7.72$ $\times10^{-3}$ & 5.1 $\times10^{-2}$\\ \hline
				$-\pi/2$& 1.28 $\times10^{-2}$ & -3.86 $\times10^{-3}$  & 7.37 $\times10^{-4}$ & 0  &8.2$\times10^{-4}$\\ \hline
				$0$& 7.31 $\times10^{-3}$  & 5.83 $\times10^{-3}$  & -1.07 $\times10^{-3}$ & 5.89 $\times10^{-3}$  &1.33$\times10^{-4}$\\ 
				\hline
				$+\pi/2$& -9.28 $\times10^{-2}$  & 9.33 $\times10^{-2}$  & -5.64 $\times10^{-2}$ & 2.59 $\times10^{-2}$ &3.9$\times10^{-4}$\\ 
				\bottomrule
		\end{tabular}}\label{tab:Table1}
	\end{table}	
	
	\begin{table}[tb]
		\centering
		\caption{\label{tab:2}Optimized amplitudes of the Fourier harmonics for different values of the desired $\Psi_1+\Psi_3$. In all the cases, $\Psi_2=\pi$. }
		\resizebox{0.48\textwidth}{!}{%
			\begin{tabular}{|c| c| c| c| c|c|} 
				\toprule
				\makecell{$\Psi_1+\Psi_3$} & \makecell{ $g_0$ } & \makecell{ $g_1$} & \makecell{ $g_2$ }& \makecell{ $g_3$ } & \makecell{ Opt. Err. } \\  \hline
				\hline
				$-\pi$&7.30 $\times10^{-3}$  & -2.78 $\times10^{-3}$ & -1.20 $\times10^{-2}$ & 0  &0.021\\ \hline
				$-\pi/2$& 8.59 $\times10^{-3}$ & 4.77 $\times10^{-3}$  & -9.40 $\times10^{-3}$ & -3.22$\times10^{-3}$  & 0.007\\ \hline
				$0$& 1.50 $\times10^{-2}$  & 1.36 $\times10^{-3}$  & 4.95 $\times10^{-3}$ & -5.81 $\times10^{-3}$  &0.045\\ 
				\hline
				$+\pi/2$& 6.89 $\times10^{-3}$  & -1.22 $\times10^{-3}$  & 2.15 $\times10^{-3}$ & -5.60 $\times10^{-4}$ &0.047\\ 
				\bottomrule
		\end{tabular}}\label{tab:Table2}
	\end{table}	
	
 \begin{figure*}[!t]
	\centering
	\includegraphics[width=0.95\linewidth]{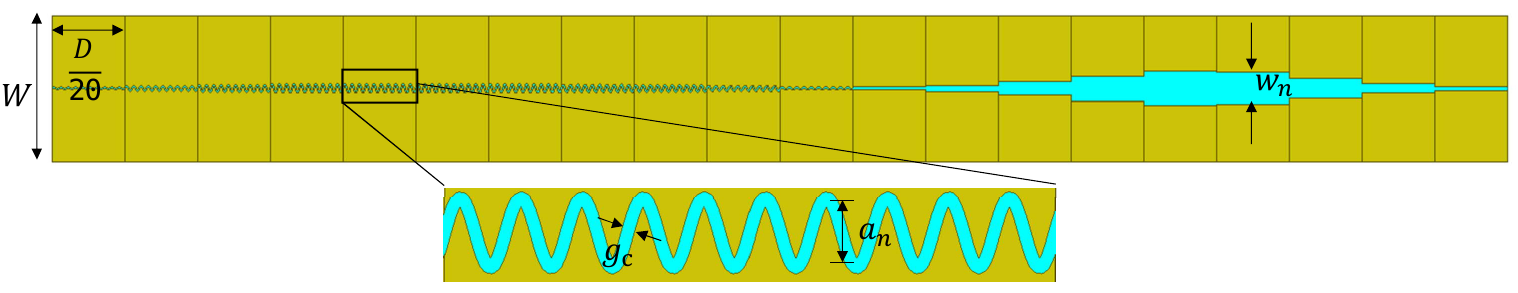}
	\caption{The unit cell of the implemented structures. The capacitive impedance of 1-11  subcells (from left to right) are controlled by meandered gaps varying with cosine functions. The gap width of the meandered slots is $g_{\rm c}=3\, \mu$m. The amplitudes of cosine functions for each subcell are different: $a_1=1.2$, $a_2=4.6$, $a_3=6.7$, $a_4=7.6$, $a_5=8.0$, $a_6=7.9$, $a_7=7.7$, $a_8=7.2$, $a_9=6.3$, $a_{10}=4.8$, $a_{11}=2.1$, with unit [$\mu$m]. The 12-20 elements are synthesized by straight gaps with the gap width $w_n$ for $n$-th subcell. $w_n$ is set to $w_{12}=5.0$, $w_{13}=11.6$, $w_{14}=25.6$, $w_{15}=48$, $w_{16}=67$, $w_{17}=62$, $w_{18}=38$, $w_{19}=78$, $w_{20}=7.4$, with unit [$\mu$m].  }\label{fig:implemetation_structure}
\end{figure*}

	The above example is only  one specific assignment of reflection phases. The reflection phases at each port can be arbitrarily defined and the corresponding surface admittance can be efficiently optimized  in MATLAB. 
	We know that if we can find the surface impedance for arbitrary $\Psi_2$ and $\Psi_{13}$, the reflection phases at all three ports can be arbitrarily engineered. 
	To check the potential of this approach, we do two parametric studies. In each study, we keep either $\Psi_2$  or $\Psi_{13}$ as constant and vary the other parameter within the range $-\pi,-\frac{\pi}{2}, 0, +\frac{\pi}{2}$. Tables~\ref{tab:Table1} and \ref{tab:Table2} present the optimized $g_m$. For each considered phase distribution,  we can always find the required surface admittance with negligible errors.

	\subsection{Practical design } \label{Sec: Practical design}

	As soon as the spatially varying surface impedance or admittance is determined, we can locally implement the surface impedance according to the impedance control method introduced in \cite{wang2018extreme}. 
	In the ideal case, if the implemented grid impedance continuously changes along the surface, the electromagnetic response of the  metasurface should be the same as analytically predicted. 
	However, in reality,  due to limited fabrication fineness, each period of the surface should be split into a finite number of subcells.
	As we demonstrated in \cite{wang2018extreme}, for the metasurfaces whose functionalities rely on evanescent modes, the number of sub-inclusions in one unit should be sufficient to accurately excite the required evanescent modes, especially for the design with strong evanescent fields excitations. 
	
		\begin{figure}[h]
		\centering
		\subfigure[]{\includegraphics[width=0.95\linewidth]{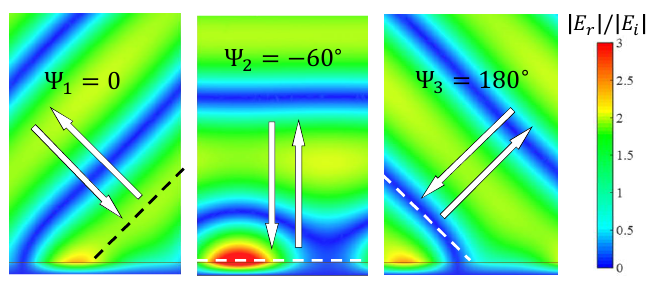}
			\label{fig:field_distribution_perfect}} 
		\subfigure[]{\includegraphics[width=0.95\linewidth]{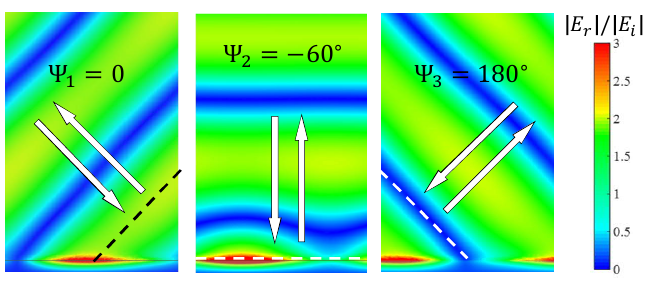}
			\label{fig: field_distribution_implementation}}
		\caption{ Simulated total fields for incidences from three ports  (a) using the impedance boundary condition and (b) particular realizations shown in Fig.~\ref{fig:implemetation_structure}. }\label{fig:comparison_boundary_implementation}
	\end{figure}

	Here, we aim to implement the design of multi-port retroreflectors with phases at each channel  $\Psi_1=0$, $\Psi_2=-\pi/3$ and $\Psi_3=\pi$. For simplicity, we choose the impedance profile of Solution~2 which corresponds to excitation of weaker evanescent modes than Solution~1. The unit cell is discretized into 20 subcells. 
	Capacitive meta-atoms are synthesized as planar capacitors with the controlled gap width $w_n$. For large capacitance, we cannot further shrink the gap due to the resolution limitation of photolithography. Instead, we use meandered gaps of a cosine shape, as shown in Fig.~\ref{fig:implemetation_structure}. By varying the amplitudes of the cosine gap of each element $a_n$, we can realize and control large sheet capacitance. After each element is engineered to ensure the required impedance value, we combine them together and  simulate the structure in HFSS without further structural optimization. Fig.~\ref{fig:field_distribution_perfect} and Fig.~\ref{fig: field_distribution_implementation} shows the simulated fields for the ideal impedance boundary  and the implemented pattern, respectively. We can see in Fig.~\ref{fig: field_distribution_implementation} that, for the actual topology, each port indeed reflects incident waves in the corresponding retrodirections.  The reflection phases and field distributions are very close to the ideal scenario simulated with the perfect impedance boundary. For incidence from Port~1, the total field is at  maximum  at the reference plane, meaning that $\Psi_1=0$.  The total field at the reference plane of Port~3 is zero which means that the reflection phase  $\Psi_3=\pi$.

	\section{Other illustrative examples }\label{Sec: other possibilities}
	
	In principle,  the scattering matrix of multi-port systems can be arbitrary defined if it does not violate fundamental physical laws. 
	For multichannel metasurfaces with more than three ports we can  use mathematical optimization to seek for proper surface impedances which ensure the defined functionalities at each channel.
	There are numerous practically important functionalities, and in this section we introduce two representative examples: multifunctional reflectors and  multichannel perfect absorbers.
	
	\subsection{Multifunctional reflectors}
	
\begin{figure}[h]
		\centering
		\subfigure[]{\includegraphics[width=0.6\linewidth]{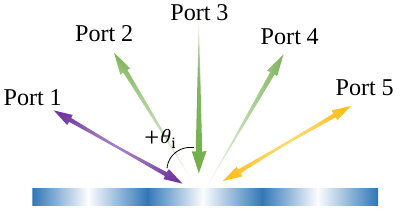}
			\label{fig:schematics_multifunctional_reflector}} 
		\subfigure[]{\includegraphics[width=0.9\linewidth]{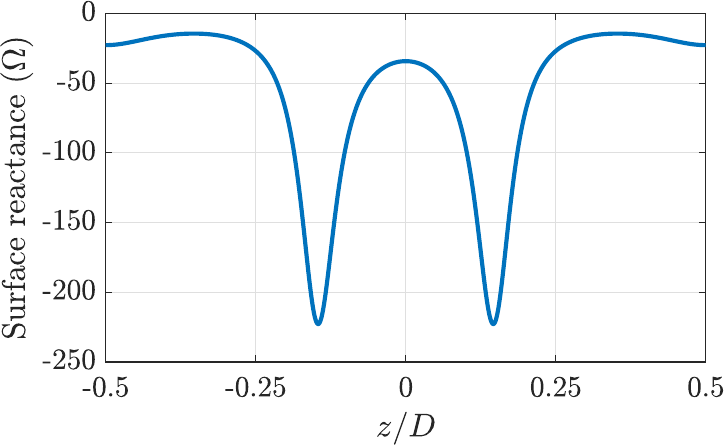}
			\label{fig: surface_reactance_multifunctional_reflector}}
		\caption{ (a) Schematic of multifunctional reflector envisioned in \cite{asadchy2017flat}.  (b) Optimized surface reactance in one period for $\theta_{\rm i}=60^\circ$.  The optimized Fourier coefficients of $Y_{\rm s}(z)$ are: $g_0=j0.037$, $g_1=-j0.013$, $g_2=-j1.25\times 10^{-4}$ and $g_3=j9.13\times 10^{-3}$. The optimized substrate thickness is $d=1.11$~mm for the defined frequency at 75~GHz and permittivity of $\epsilon_{\rm d}=4.2$. The incidence is assumed to be TE-polarized.  }\label{fig:multifunctional_reflectors}
	\end{figure}

	Here, our target is to find the surface impedance realizing  the multifunctional reflector envisioned in \cite{asadchy2017flat}.  
	The device is a five-channel metasurface with periodicity $D=2\lambda/\sin\theta_{\rm i}$. As shown in Fig.~\ref{fig:schematics_multifunctional_reflector}, the metasurface can act as a perfect retroreflector when illuminated from $\theta=\pm\theta_{\rm i}$, while under the normal incidence, it equally splits the beam into directions of $\theta=\pm \arcsin(\frac{\sin\theta_{\rm i}}{2})$.
	In this case, we need to control the scattering properties from three ports: for incidence from Port~3 the wave equally splits between Ports~2 and 4, so that $|S_{43}|=|S_{23}|=0.707$; for incidence from Port~1 ($\theta=+\theta_{\rm i}$) or Port~5 ($\theta=-\theta_{\rm i}$), the waves bounce back to the corresponding retrodirections with $|S_{11}|=|S_{55}|=1$. 
	
\begin{figure}[h]
		\centering
		\includegraphics[width=0.9\linewidth]{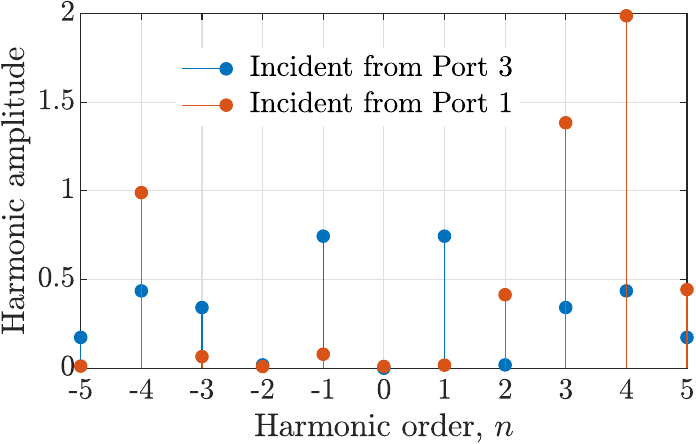}
		\caption{Amplitudes of scattered harmonics for incidence from Port 1 ($\theta=+60^\circ$) and Port 3 ($\theta=0^\circ$).  }\label{fig:scattering_harmonics_multifunctional_reflectors}
	\end{figure}

In the optimization, we introduce four unknowns $g_m$ ($m=0,\pm1, \pm 2, \pm 3$), and also set a constraint in the code to ensure a purely capacitive surface impedance (for simplicity of realization).
	In addition, the substrate thickness is set as another unknown in order to increase the optimization freedom. 
	The optimized surface impedance for $\theta_{\rm i}=60^\circ$ is shown in the caption of Fig.~\ref{fig:multifunctional_reflectors}. The scattered harmonics for incidences from Port 3 (beam splitter operation) and Port 1 (performs as a retroreflector) are  calculated  in Fig.~\ref{fig:scattering_harmonics_multifunctional_reflectors}.   
	We can see that for the normal incidence, the surface splits the impinging energy equally between Port~2 ($n=-1$) and  Port~4 ($n=+1$), while Port~1 ($n=-2$) and Port~5 ($n=+2$) receive zero energy. 
	For incidence from Port~1 ($\theta=+60^\circ$), all the energy is reflected back in the retrodirection, corresponding to harmonic $n=-4$.
	Since we look for the surface impedance as an even function with respect to the $z$-axis,  the scattered harmonics for incidence from Port~5 are the same as for incidence from Port~1.
	
\subsection{Multichannel perfect absorbers}	

Following the generally accepted terminology \cite{landy2008perfect,Alaee_2017}, we define a planar surface as a ``perfect absorber'' if the theoretical absorptivity equals unity (no reflection and no transmission) for plane-wave illumination at a certain (single) frequency, for a certain polarization, and a certain incident angle. It is commonly believed that, in passive metasurfaces, such perfect absorber can be designed only for  one specific incident angle, with a few exceptions based on the use of spatial dispersion, e.g. \cite{zhirihin2017mushroom}.
Deviating from the defined angle, absorptance inevitably decreases due to impedance mismatch between the incident wave and the metasurface structure \cite{diem2009wide}. Full absorption for multiple illumination angles can be realized by employing some tunable materials in the absorbing structure which can instantaneously adjust the input impedance of metasurface with the change of the incident angles \cite{wang1803graphene}.
Although these devices can absorb energy from more than one direction, their operations rely on the modification of material properties and therefore they cannot work for simultaneous illuminations from multiple directions.  Physical  realizations of all-angle perfect absorbers (perfectly matched layers, PML) \cite{Ziolk1,Ziolk2,Active_PML} require the use of active elements \cite{Active_PML} or engineered strong spatial dispersion in bulk metamaterials \cite{PML1998}. To the best of our knowledge, no experimental realizations of such devices have been demonstrated. 

\begin{figure}[h]
	\centering
	\subfigure[]{\includegraphics[width=0.9\linewidth]{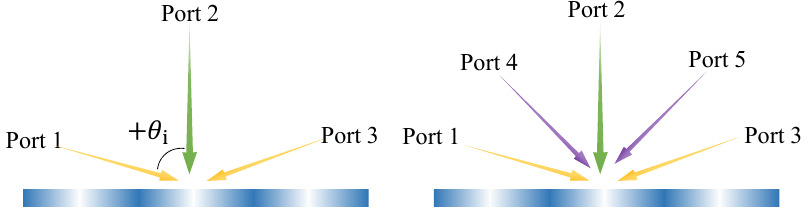}
		\label{fig:schematics_multichannel_absorbers}} 
	\subfigure[]{\includegraphics[width=0.9\linewidth]{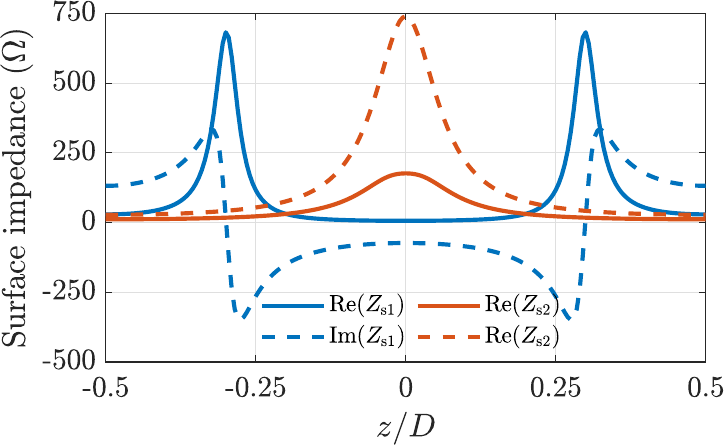}
		\label{fig: surface_impedance_multichannel_absorber}}
	\caption{ (a) Schematics of three- (left) and five- (right) ports multichannel absorbers.  (b) Optimized surface impedance (in one period) for perfect absorption three- and five- ports metasurface absorbers.  The period is chosen as $D=0.419\lambda$. The optimized parameters for $Z_{\rm s1}(z)$ reads: $g_0=(1.4+3.16j)\times10^{-3}$,  $g_1=g_{-1}=(-1.08+51.9j)\times10^{-4}$,
    $d=0.5$~mm and $\epsilon_{\rm d}=4.2$.
	For $Z_{\rm s2}(z)$, it reads: $g_0=(6.5-15.7j)\times10^{-3}$,  $g_1=g_{-1}=(-3.1+7.2j)\times10^{-3}$, $d=1.0$, and $\epsilon_{\rm d}=1.4$.   The incident wave is TM-polarized. }\label{fig:multichannel_absorber}
\end{figure}

Here, we  overcome this problem by using gradient metasurfaces with controlled channel responses. We utilize the same approach as presented above, introducing different sets of evanescent modes at different incident angles. These evanescent modes provide enough freedom to find the boundary condition which ensures full absorption for multiple incidence scenarios. 
We set the period of the metasurface below the diffraction limit, $D<\lambda/2$, so that there are no higher-order diffraction modes produced by the metasurface illuminated by plane waves from any direction.
As illustrated in the left of Fig.~\ref{fig:schematics_multichannel_absorbers}, as the simplest example, we first target to realization of perfect absorption for two different incidence directions, $\theta=0^\circ$ (Port 2) and $\theta=+75^\circ$ (Port 1), and perfect absorption at $\theta=-75^\circ$ (Port 3) is unconsciously ensured due to reciprocity.
The surface admittance is assumed to be a complex value varying as an even function of $z$, $Y_{\rm s}(z)=Y_{\rm s}(-z)$. Therefore, the Fourier coefficients of $Y_{\rm s}(z)$  satisfy $g_m=g_{-m}$. 
\begin{figure}[t]
	\centering
	\includegraphics[width=0.9\linewidth]{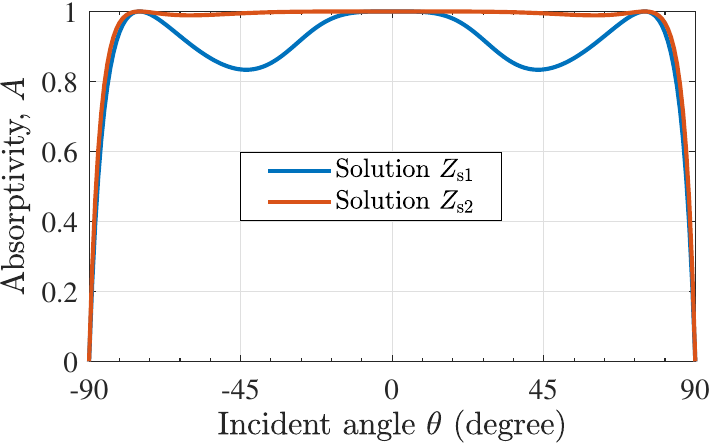}
	\caption{ Absorptivity of the optimized impedance sheets in Fig.~\ref{fig: surface_impedance_multichannel_absorber} as a function of the incident angle.  }\label{fig:absorptivity_multichannel_absorber}
\end{figure}
Here, we use only two unknowns ($g_0$ and $g_{\pm 1}$) in the optimization. Note that the optimal sets of device parameters are not unique, and here we present one of the optimized solutions $Z_{\rm s1}$ ($Z_{\rm s1}=1/Y_{\rm s1}$), shown as the blue curves in Fig.~\ref{fig: surface_impedance_multichannel_absorber}. The absorptivity of the optimized surface as a function of the incident angle are calculated in Fig.~\ref{fig:absorptivity_multichannel_absorber} (blue curve). Due to the enforcement of perfect absorption at $\theta=0^\circ$ and  $\theta=\pm75^\circ$, the absorptivity in the angle spectrum remains high (above 80\% between $\theta=-83^\circ$ and $\theta=+83^\circ$). 
However, there exist obvious absorption dips (close to $\theta=\pm 45^\circ$) between the two defined angles of perfect absorption.  

\begin{figure}[t!]
	\centering
	\includegraphics[width=0.9\linewidth]{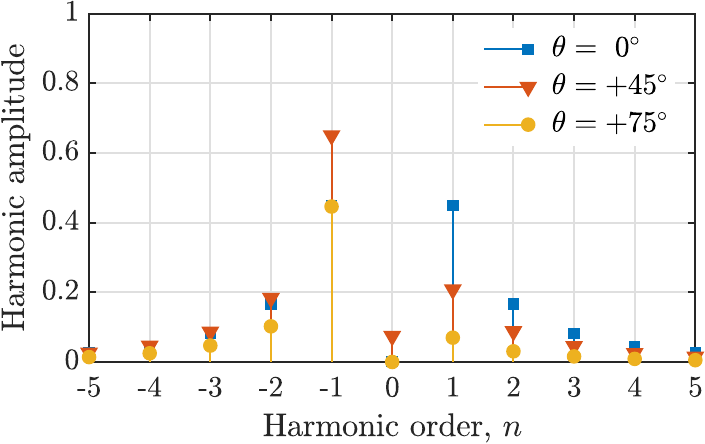}
	\caption{Amplitudes of scattered harmonics of five-port perfect absorbers for incidences from Port 2 ($\theta=0^\circ$), Port~4 ($\theta=+45^\circ$) and Port~1 ($\theta=+75^\circ$).   }\label{fig:scattering_harmonics_multichannel_absorber}
\end{figure}

To avoid performance deterioration between the two perfect-absorption angles, one can impose more requirements on the metasurface performance, defining additional angles of perfect absorption.  Here, we add one more objective in the optimization, enforcing perfect absorption at $\theta=\pm45^\circ$, as shown in the right of \ref{fig:schematics_multichannel_absorbers}(five-port perfect absorber).
The optimal surface impedance for five-port perfect absorber is shown in \ref{fig: surface_impedance_multichannel_absorber} (red curves). One can see from Fig.~\ref{fig:absorptivity_multichannel_absorber}(red curve) that, after ensuring full absorption at $\theta=\pm45^\circ$, the device acts as a near-perfect absorber ($A>99\%$) at all angles between $\theta=-80^\circ$ and $\theta=+80^\circ$.  
Figure~\ref{fig:scattering_harmonics_multichannel_absorber} shows the excited Floquet harmonics of the five-port perfect absorber for incidences from different directions. As expected, different sets of  evanescent modes are generated at the three incident angles. 
The implementation of such impedance sheet with both resistance and reactance parts can be done using the impedance control methods introduced in \cite{wang2018extreme}.

From the physical point of view, this structure realizes multiple-angle perfect absorption by engineering spatial dispersion in the metasurface, resulting in wide-angle metasurface absorbers. In contrast to earlier studies \cite{PML1998} which considered bulk metamaterials, the desired performance is achieved by proper structuring of a single material sheet, allowing simple practical realizations.

\section{Conclusions}
	
	In this paper, we have introduced a general and efficient method for independent control of port responses in multichannel multifunctional metasurfaces. To implement a desired scattering matrix of metasurface,
	our idea is to find a set of evanescent modes excited at each incidence scenario, which together with the defined propagating modes  simultaneously satisfy one impedance boundary condition. 
	Instead of looking for a purely analytical solution of the problem, we use mathematical optimization tools to find the Fourier coefficients of the surface admittance expansion which realize the defined scattering matrix. 
	
From the physical point of view, the proposed method allows engineering of strong spatial dispersion (desired non-local response of the metasurface is defined by a set of evanescent modes carrying power along the metasurface), going beyond the known designs which are limited to engineer electric, magnetic, and bianisotropic properties of metasurfaces \cite{bian}. From the applications point of view, the presented examples demonstrate possibilities of creation of multichannel metasurface devices for rather general control of reflection and transmission of waves between multiple specified directions in space. In particular, designs of phase-controlled multichannel retroreflectors, multifunctional reflectors, and perfect absorbers for multiple incident angles have been presented as examples. Importantly, the proposed multichannel functional reflectors are realized by structuring only one planar material sheet, without a need to engineer volumetric artificial materials.

	\section*{Acknowledgments}
	
	This work received funding from the European
	Union’s Horizon 2020 research and innovation programme -- Future Emerging Topics (FETOPEN) under grant agreement No 736876. The work is also supported by Aalto ELEC Doctoral school. The authors would thank Dr. Viktar Asadchy for helpful discussions.

\bibliography{references}

\end{document}